\documentclass[12pt,preprint]{aastex}
\usepackage{booktabs}

\newcommand{\lsim}{\raise0.3ex\hbox{$<$}\kern-0.75em{\lower0.65ex\hbox{$\sim$}}}
\newcommand{\gsim}{\raise0.3ex\hbox{$>$}\kern-0.75em{\lower0.65ex\hbox{$\sim$}}}
\newcommand{\propsim}{\raise0.3ex\hbox{$\propto$}\kern-0.75em{\lower0.65ex\hbox{$\sim$}}}

\begin{document}
    
\title{The initial evolution of SN\,2011dh: The importance of inhomogeneities}

\author{C.-I. Bj\"ornsson\altaffilmark{1}}
\altaffiltext{1}{Department of Astronomy, AlbaNova University Center, Stockholm University, SE--106~91 Stockholm, Sweden.}
\email{bjornsson@astro.su.se}

\begin{abstract}
SN\,2011dh is rather unique in that it offered detailed observations of the initial phase in the radio as well as optical regimes. This makes possible a comparison between models used to deduce properties of the outer envelope of the supernova ejecta. It is shown that a consistent description suggests the forward shock to have started in the piston phase with constant velocity, and only later, around 50\,days, transitioned to the standard model, which is independent of initial conditions. In addition, observations imply that the radio source is inhomogeneous with a covering factor $\approx$ 50\%.

It is emphasised that the deduced properties of the synchrotron source are very sensitive to the presence of inhomogeneities; for example, a covering factor of 50\% increases the ratio of the energy densities of relativistic electrons and magnetic field by several orders of magnitude as compared to a homogeneous source. The shallow density gradient in the envelope causes substantial deceleration of the forward shock. This is used to argue that the magnetic field scales with radius as $B\propto 1/R$ rather than with time as $B\propto 1/t$; this is similar to SN\,1993J. Attention is also drawn to the similarities between the flat spectra of compact, extragalatic radio sources and the evolution of radio supernovae; e.g., the scaling of the magnetic field and the constant brightness temperature.

\end{abstract}


\section{Introduction}
An understanding of the structure and velocity of the outer envelope of the supernova ejecta is important in order to connect the various types of supernovae to their progenitor stars \cite[see][for a review of the observed characteristics of the various types of supernovae]{fil97}. Although direct identification of the progenitor star is now starting to become available \cite[e.g.,][]{sma15,van23}, their association still relies, to a large extent, on theoretical considerations. The properties of the outer envelope are sensitive both to the mass and evolutionary history of the progenitor. Here, a central question is the extent of mass-loss during the lifetime of the progenitor and, in particular, whether it is due to stellar winds or if mass-transfer in a binary system is needed. It is thought that the mass-loss increases from the standard type II to the stripped envelope Ib/c supernovae. Type IIb supernovae constitute an interesting transition class, since they have retained s small amount of hydrogen in an extended envelope. 

In order to probe these outermost parts, observations need to be done early on; for example, observations of the shock break-out and the cooling radiation from behind the shock have made it possible to constrain the properties of the progenitor \citep{ber18,far25a}. Another way to learn about the ejecta structure is to observe the radiation coming from its interaction with the circumstellar medium. In addition, since the latter is likely due to a stellar wind, it gives information of the properties of the progenitor prior to explosion. 

The evolution of the shockwaves produced by this interaction is best followed in the radio; in particular, when the radio emission region is spatially resolved through VLBI-observations, for example, SN\,1993J \citep{bar02,mar09}. In order to connect the properties of the circumstellar medium to those of the ejecta, a model is needed. The standard one used for this purpose is homogeneous and spherically symmetric. Furthermore, an important assumption is made in that the initial conditions for the onset of the interaction no longer affect the evolution; this makes the evolution self-similar \citep{che82a}. 

The conclusions drawn from observations are no more secure than the model used to deduce them; hence, it is crucial to evaluate the consistency of the results obtained from a given model. It was argued in \cite{bjo25} that the standard model does not give consistent results for SN\,1993J. Since this is one of the most well-observed supernovae, it could be shown that these deviations were likely due to the omission of the initial phase, which is characterised by a constant velocity of the forward shock.

Although SN\,2011dh was not as well-observed as SN\,1993J in the radio, it was rather unique in that it had early optical spectral observations \citep{erg14,mar14} during  the time when the first radio observations were made \citep{sod12, kra12,hor13}. This makes possible a comparison of the ejecta properties deduced from two different models; namely, the model used to analyse the optical spectra and the standard model for the radio observations. It is noteworthy that these two independent ways to constrain the ejecta properties gave very different results: (1) The absorption lines in the optical indicated a shallow density structure for the ejecta, while the radio observations required a very steep one. (2) The optical spectra showed ejecta velocities substantially higher than obtain from the standard model in the radio regime. 

Another important result of the observations is the inferred properties of the synchrotron emitting plasma. It has been argued that the shocked region is often inhomogeneous and that this can severely bias the conclusions drawn from observations, in particular, regarding the strength of the magnetic field and the energy density of the relativistic electrons \citep{bjo24}. As discussed in \cite{bjo15}, the evolution of the radio emission in SN\,1993J indicated a value for the magnetic field, which was substantially below that obtained from the standard model \citep{f/b98}. The fact that the optical observations of SN\,2011dh showed ejecta velocities larger than those resulting from the standard model, directly suggests that the synchrotron emission region is inhomogeneous; i.e., the emitting surface of the radio emission region is smaller than  that of the actual source.

The present paper has two aims: (1) To reconcile the diverging conclusions regarding the density structure of the outer envelope in SN\,2011dh. It is argued that the almost constant velocity deduced from the radio observations is not due to a steep density slope of the outer envelope but is, instead, a result of the initial piston phase \citep{h/s84}. It is also shown that the observations can be well accounted for by a transition from this initial phase to the standard model as described by the approximate analytical solution derived by \cite{t/m99}. (2) Estimating the covering factor of the radio emission region from the different velocities derived from the optical and radio regimes. Furthermore, it is shown that the deduced properties of the synchrotron plasma are very sensitive to inhomogeneities; for example, even a factor two decrease of the covering factor causes more than two orders of magnitude increase in the ratio between the energy densities in relativistic electrons and magnetic field.

A comparison between a homogeneous and an inhomogeneous source is done in Section \ref{sect2}; in particular, it is shown how the description of a homogeneous source can be modified by the introduction of a covering factor to account also for an inhomogeneous source. The VLBI-observations of SN\,1993J are used as an example to show how a value for the covering factor can be estimated. The observations of SN\,2011dh are discussed in Section \ref{sect3}. The analytical solution derived by \cite{t/m99} is used to constrain the evolution of the forward shock in Section \ref{sect3a}; of particular importance here is the spatially resolved VLBI-observation on day 453 by \cite{dew16}. The evolution of the synchrotron self-absorption frequency is considered in Section \ref{sect3c}. It is emphasised that in an inhomogeneous source, the evolution of the spectral flux depends on the covering factor, while the self-absorption frequency does not. This is used in Section \ref{sect3d} to argue that the decline in the peak spectral flux starting at around day 100, is due to a decreasing covering factor. The implications of the observations of SN\,2011dh are discussed Section \ref{sect4}. Furthermore, the similarities between SN\,2011dh and SN\,1993J are emphasised. 
 The conclusions of the paper are collected in Section\,\ref{sect5}. The notation follows closely that used in \cite{bjo25} and \cite{t/m99}. Numerical results are mostly given using cgs-units. When this is the case, the units are not written out explicitly.

\section{An inhomogeneous versus a homogeneous source model}\label{sect2}
The standard synchrotron source model is homogeneous and spherically symmetric. It is normally assumed that the radiating electrons have a distribution of Lorentz factors ($\gamma$) according to  $n(\gamma) = K_{\rm o} \gamma^{-{\rm p}}$ for $\gamma > \gamma_{\rm min}$ and $\rm {p\,>\,2}$. As discussed in \cite{bjo24}, this leads to a closure relation for the radius ($R$), magnetic field ($B$) and a parameter $y$ given by
\begin{equation} 	
	y = (\rm p-2) \gamma_{\rm min}^{\rm p-2}\frac{U_{\rm rel}}{U_{\rm B}}\frac{R_{||}}{R}.
	\label{eq2.1}
\end{equation}
Here, $U_{\rm rel}$ and $U_{\rm B}$ are the energy densities in relativistic electrons and magnetic fields, respectively, and $R_ {||}$ is the average line of sight extension of the source. It is seen that $y$ combines the parameters describing much of the unknown physics in the source. The upper limit in the electron distribution ($\gamma_{\rm max}$) is sometimes important, for example, when p$\,<\,$2. When this is the case, a factor $1/[1-(\gamma_{\rm max}/\gamma_{\rm min})^{\rm 2-p}]$ should be added to the RHS in Equation (\ref{eq2.1}). For p$\,=\,$2, this factor becomes $1/\{(\rm p-2)\ln(\gamma_{\rm max}/\gamma_{\rm min})\}$. One may note that the unknown physics also includes the value of p. It is often assumed that first-order Fermi-acceleration is the mechanism, which injects the relativistic electrons at the forward shock; this implies p$\,=\,$2. However, in supernovae, observations indicate somewhat higher values $(\rm 2\,\lsim\,p\,\lsim\,3)$.
 
In order to make use of this closure relation, three independent observables are needed. The most straightforward ones to obtain are  $\nu_{\rm abs}$, the frequency where the spectral flux peaks, and $F_{\nu_{\rm abs}}$, which is the corresponding spectral flux. The emission from the supernova proper can be inverse Compton scattered by the relativistic electrons. This scattered radiation is sometimes observed in the X-ray regime. When this is the case, the ratio of these two luminosities provides a third observable. The consequences of this was discussed in \cite{bjo24}. In the presented paper, the third observable is, instead,  the radius obtained directly from spatially resolved VLBI-observations.

\subsection{The homogeneous source model}\label{sect2a}
\begin{figure}[h]
	\includegraphics[width=\linewidth]{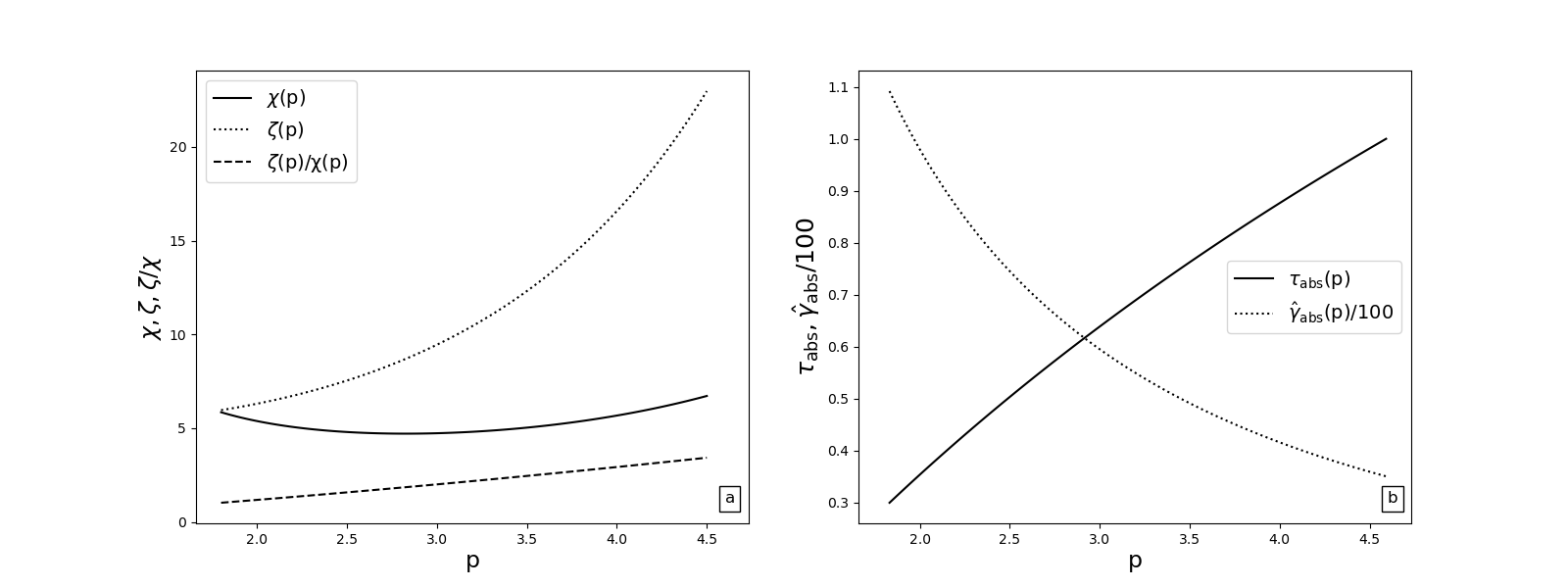}
	\caption{The dependence of basic synchrotron properties on p as derived in the Appendix. (a) The emissivity ($\chi(\rm p)$) and absorptivity ($\zeta(\rm p)$). (b) The emission averaged Lorentz factor of the electrons radiating at the spectral peak ($\hat{\gamma}_{\rm abs}(\rm p)$) and the corresponding optical 		depth ($\tau_{\rm abs}(\rm p)$)}
	\label{fig1}
\end{figure}

\begin{figure}[h]
	\includegraphics[width=\linewidth]{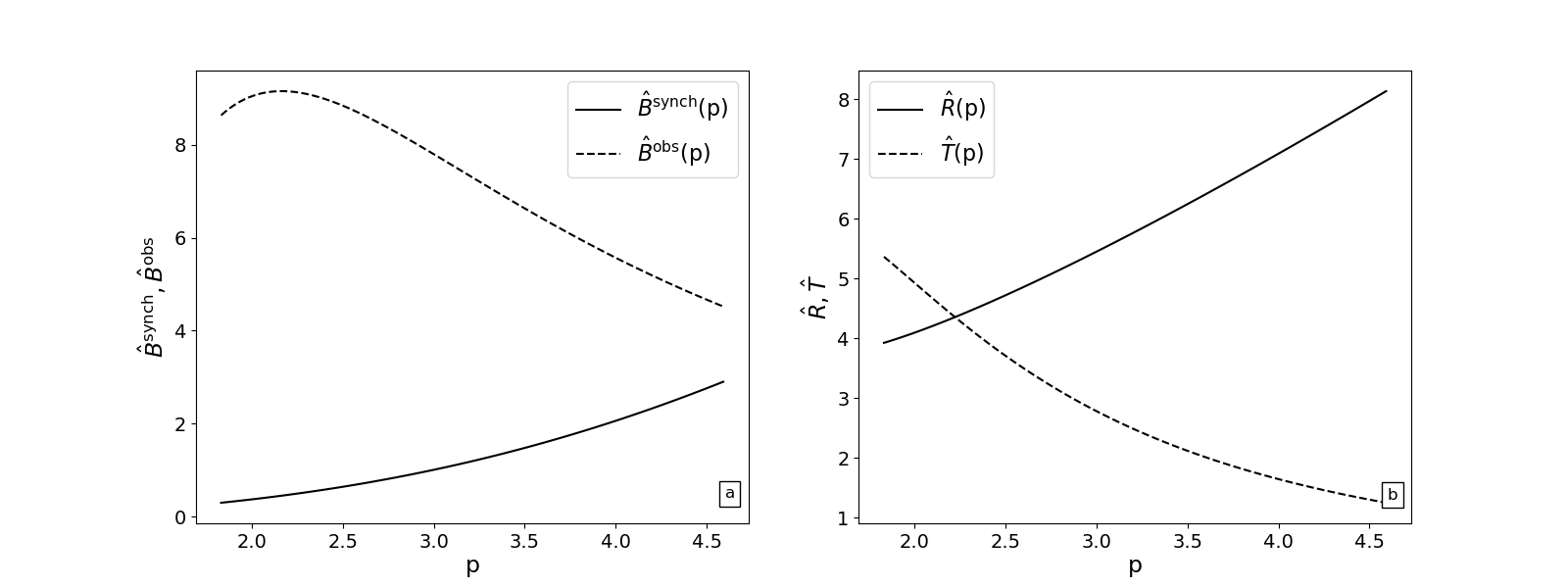}
	\caption{The dependence of deduced properties of a synchrotron source on p as derived in the Appendix. (a) The strength of the magnetic field measured 	in two ways: $B^{\rm synch}(\rm p)$ corresponds to the standard procedure, while $B^{\rm obs}(\rm p)$ relies on the distance independent value of the        	observed brightness temperature. b) The normalisation of the radius of the source ($\hat{R}(\rm p)$) and its brightness temperature ($\hat{T}(\rm p)$)}
	\label{fig2}
\end{figure}

It is useful to define a quantity $\gamma_{\rm abs} \equiv (\nu_{\rm abs}/\nu_{\rm B})^{1/2}$, where $\nu_{\rm B}\,(=eB/2\pi mc)$ is the cyclotron frequency. $\gamma_{\rm abs}$ corresponds to the emission averaged Lorentz factor of the electrons radiating at $\nu_{\rm abs}$. It is shown in the Appendix that
\begin{equation}
	\gamma_{\rm abs} = \hat{\gamma}_{\rm abs}({\rm p}) y^{2/(\rm{2p+13})} F_{\nu_{\rm abs},27}^{1/({\rm 2p +13})},
	\label{eq2.2}
\end{equation}
where $F_{\nu_{\rm abs},27} \equiv F_{\nu_{\rm abs}}/10^{27}$ and the function $\hat{\gamma}_{\rm abs}({\rm p})$ is shown in Figure \ref{fig1} together with the underlying emissivity ($\chi(\rm p)$) and absorptivity ($\zeta(\rm p)$). Hence, the strength of the magnetic field can be written
\begin{equation}
	B = \hat{B}^{\rm synch}({\rm p}) \frac{\nu_{\rm abs,10}}{y^{4/(\rm{2p+13})} F_{\nu_{\rm abs},27}^{2/({\rm 2p +13})}},
	\label{eq2.3}
\end{equation}
where $\nu_{\rm abs,10} \equiv \nu_{\rm abs}/10^{10}$ and the function $\hat{B}^{\rm synch}({\rm p})$ is shown in Figure \ref{fig2}a. One may note that the value of $\hat{\gamma}_{\rm abs}(\rm p)$ decreases with an increasing value for p. The  reason is that the spectral flux then receives an increasing relative contribution from lower energy electrons. This leads to a decrease in the emission averaged Lorentz factor of the electrons radiating at a given frequency. 

Likewise, the standard expression for the source radius can be written
\begin{equation}
	R_{15} = \hat{R}({\rm p}) \frac{F_{\nu_{\rm abs},27}^{({\rm p+6})/({\rm 2p +13})}}{\nu_{\rm abs,10}\,y^{1/({\rm 2p + 13})}},
	\label{eq2.4}
\end{equation}
where $R_{15} \equiv R/10^{15}$ and the function $\hat{R}({\rm p})$ is shown in Figure \ref{fig2}b.
Instead of the expression for the radius in Equation (\ref{eq2.4}), it is sometimes convenient to use the brightness temperature, in particular, since it is independent of distance to the source. With $T_{\nu}^{\rm br} = (1/8\pi^2 k) F_{\nu}(c/R\nu)^2$, the observed brightness temperature is given by
\begin{equation}
	T_{\nu,10}^{\rm br,obs} = 4.4\times 10^{-4 }\frac{f_{\nu}}{\rm mJy} \frac{1}{\nu_{10}^2\,\theta_{\rm mas}^2},
	\label{eq2.5}
\end{equation}
where $T_{\nu,10}^{\rm br,obs} \equiv T_{\nu}^{\rm br,obs}/10^{10}$,  $f_{\nu}$ is the observed spectral flux in units of mJy and $\theta_{\rm mas}$ is the angular radius in units of mas. Furthermore, since the source is assumed to be spherically symmetric, it subtends a solid angle  $\pi \theta^2$.

The brightness temperature in the optically thick part of the spectrum is a direct measure  of the average energy of the particles radiating at a given frequency. For synchrotron radiation, this implies $T_{\nu_{\rm abs}}^{\rm br, synch} \propto \gamma_{\rm abs}$ at $\nu = \nu_{\rm abs}$ and, for example, that $B \propto    (T_{\nu_{\rm abs}}^{\rm br,synch})^{-2}$. The expression for $T_{\nu_{\rm abs}}^{\rm br,synch}$ is given by (see Appendix)
\begin{equation}
	T_{\nu_{\rm abs},10}^{\rm br,synch}= \hat{T}({\rm p}) y^{2/(\rm{2p+13})} F_{\nu_{\rm abs},27}^{1/({\rm 2p +13})},
	\label{eq2.6}
\end{equation}
where $T_{\nu_{\rm abs},10}^{\rm br,synch} \equiv T_{\nu_{\rm abs}}^{\rm br,synch}/10^{10}$ and the function $ \hat{T}({\rm p})$ is shown in Figure \ref{fig2}b. It is then seen from Equations (\ref{eq2.5}) and (\ref{eq2.6}) that equating $ T_{\nu_{\rm abs}}^{\rm br,obs} =  T_{\nu_{\rm abs}}^{\rm br,synch}$ is equivalent to Equation (\ref{eq2.4}). One may also note that this allows an alternative to Equation (\ref{eq2.3}) 
\begin{equation}
	B = \hat{B}^{\rm obs}(\rm p)\frac {\nu_{\rm abs,10}}{\left(T_{\nu_{\rm abs},10}^{\rm br,obs}\right)^2},
	\label{eq2.3a}
\end{equation}
where ${\hat{B}^{\rm obs}(\rm p)}$ is shown in Figure \ref{fig2}a.

\subsection{The inhomogeneous source model}\label{sect2b}
The description of an inhomogeneous source was discussed in \cite{bjo24}. It was argued that the essential features could be accounted for by a model where the inhomogeneities were approximated as a superposition of individual subcomponents, each of which modelled as a standard homogeneous source; for example, the flat-topped spectra sometimes observed would be due to a range of  $\nu_{\rm abs}$-values reflecting the different properties of the individual subcomponents. Each of the subcomponents has a radius $R_{\rm sub}$ and a corresponding covering factor $\eta_{\rm cov} \equiv (R_{\rm sub}/R)^2$, where R is the actual radius of the source. One should note that a spatially resolved, inhomogeneous source would appear homogeneous, if the scale of the inhomogeneities is below the spatial resolution of the observations.

The results in Section \ref{sect2a} can then be used also for an inhomogeneous source with $F_{\nu_{\rm abs}}$ substituted by $F_{\nu_{\rm abs}}/\eta_{\rm cov}$.  As a result of this substitution, $F_{\nu_{\rm abs}}$ now corresponds to a given subcomponent, while $R$ still refers to the radius of the whole source. One then finds from Equation (\ref{eq2.4})
\begin{equation}
	R_{15} = \hat{R}({\rm p}) \frac{F_{\nu_{\rm abs},27}^{({\rm p+6})/({\rm 2p +13})}}{\nu_{\rm abs,10}\,(y\eta_{\rm cov}^{\rm p+6})^{1/({\rm 2p + 13})}}.
	\label{eq2.7}
\end{equation}
In the case of an inhomogeneous source, Equation (\ref{eq2.7}) shows that an observation of $R$ together with $F_{\nu_{\rm abs}}$ and $\nu_{\rm abs}$ determines the combination $y \eta^{\rm p+6}_{\rm cov}$. The same result is obtained from equating the observed brightness temperature in Equation (\ref{eq2.5}) (with  $f_{\nu_{\rm abs}} \rightarrow f_{\nu_{\rm abs}}/\eta_{\rm cov}$ ) to the synchrotron value in Equation (\ref{eq2.6}).
Likewise, the value for the magnetic field in the subcomponent can be written
\begin{equation}
	B = \hat{B}^{\rm synch}({\rm p}) \frac{\nu_{\rm abs,10}}{F_{\nu_{\rm abs},27}^{2/({\rm 2p +13})}}\left(\frac{\eta_{\rm cov}}{y^2}\right)^{2/(\rm 2p+13)}.
	\label{eq2.8}
\end{equation}
or, alternatively,
\begin{equation}
	B = \hat{B}^{\rm obs}(\rm p) \left(\frac{\eta_{\rm cov}}{T_{\nu_{\rm abs},10}^{\rm br,obs}}\right)^2.
	\label{eq2.8a}
\end{equation}

There are a few things to note from Equations ({\ref{eq2.7}) and (\ref{eq2.8a}). It is seen from Equation (\ref{eq2.8a}) that the deduced strength of the magnetic field is quite sensitive to the covering factor ($B\propto \eta_{\rm cov}^2$). The reason is that a lower value for $\eta_{\rm cov}$ implies a higher brightness temperature in the subcomponent and, hence, a larger value for $\gamma_{\rm abs}$, which, in turn, decreases the value of $B$.
Furthermore, since observations of an inhomogeneous source result only in a value for $y \eta^{\rm p+6}_{\rm cov}$ (see Equation \ref{eq2.7}), the deduced internal properties of the subcomponent are very sensitive to the covering factor, i.e., $y\propto 1/\eta_{\rm cov}^{\rm p+6}$.

In conclusion, assuming a homogeneous source (i.e., $\eta_{\rm cov} = 1$) when, in fact, $\eta_{\rm cov} < 1$, will result in a much smaller, artificial value of $y$, which, in turn, implies an artificially large value of $B$. This also erroneously underestimates the value of the energy density of relativistic electrons, since $U_{\rm rel}/U_{\rm B} \propto y$ (see Equation \ref{eq2.1}). 

\subsection{SN\,1993J}\label{sect2c}
SN\,1993J is one of the most extensively observed radio supernovae. It was spatially resolved by VLBI-observations from an early date \citep{bar02,mar09}. Although spectra and light curves could be rather well accounted for with a homogeneous model after a few hundred days \citep{wei07}, there are clear observational indications that its source structure was inhomogeneous \citep{bie03,mar24,far25b}. Therefore, it is instructive to compare the parameter values resulting from an assumption of homogeneity to those when the inhomogeneities are accounted for. Although the initial evolution is unusual, after, roughly, a hundred days, it transitions to a phase in which the emission is rather well fitted by one dominant subcomponent. The evolution of this component is characterised by a roughly constant value of $F_{\nu_{\rm abs}}$ and $\nu_{\rm abs} \propto R^{-1}$ \citep{f/b98}. This implies that the observed brightness temperature is constant (see Equation \ref{eq2.5}), which, in turn from Section \ref{sect2a}, suggests $\gamma_{\rm abs} =$ constant and $B\propto R^{-1}$.

In most situations, the peak of the light curve for a given frequency $\nu$ does not occur at the  time when this frequency corresponds to the spectral peak. However, as shown in \cite{bjo22}, when $F_{\nu_{\rm abs}}$ is constant and $\nu_{\rm abs} \propto R^{-1}$, these two times coincide (see also Section \ref{sect3d}). Hence, for SN\,1993J, $F_{\nu_{\rm abs}}$ and $\nu_{\rm abs}$ can be directly obtained from the light curves. The best observed light curve in \cite{wei07} is the one corresponding to $\lambda = 6$\,cm. The peak of this light curve is estimated to be at $t = 143$\,days, which then yields $f_{\nu_{\rm abs}} = 115$\,mJy for $\nu_{\rm abs,10} = 0.5$. Furthermore, at this time, the average of the angular radii measured by \cite{bar02} at the two frequencies $\nu = 8.4$\,GHz and $\nu = 14.8$\,GHz is $\theta = 0.38$\,mas. It is then found from Equation (\ref{eq2.5}) that $T^{\rm br,obs}_{\nu_{\rm abs}, 10} = 1.4/\eta_{\rm cov}$.

Spectral modelling of SN\,1993J by \cite{f/b98} indicated p=2.7 in a non-cooling scenario. The expected brightness temperature from synchrotron radiation is obtained from Equation (\ref{eq2.6}) as $T^{\rm br,synch}_{\nu_{\rm abs}, 10} = 3.3 \,y^{1/9.2} /\eta_{\rm cov}^{1/18.4}$, where $\hat{T} (2.7) = 3.2$ (see Figure \ref{fig2}b) has been used. Furthermore, the distance to SN\,1993J was taken to be 3.63\,Mpc so that $F_{\nu_{\rm abs},27} = 1.8$. Equating these two brightness temperatures yields $y\eta_{\rm cov}^{8.7} = 3.8\times 10^{-4}$. 

This shows explicitly the sensitivity of the deduced properties of the synchrotron emission region (i.e., $y$) to the covering factor (i.e., $\eta_{\rm cov}$). \cite{f/b98} assumed $\eta_{\rm cov} = 1$, which implies $y = 3.8 \times 10^{-4}$. This results in a very small value for $U_{\rm rel}/U_{\rm B}$ (see Equation \ref{eq2.1}); in fact, the value of $B$ is so large that synchrotron cooling becomes important (i.e., $R_{||}/R \ll 1$). As discussed in \cite{bjo15}, such a high value for the magnetic field strength has several unwanted consequence; for example, it implies a very high value for the thermal energy density behind the forward shock. This, in turn, requires a kinetic energy in the envelope of the expanding ejecta ($\approx 4.4 \times 10^{51}$), which is at least an order of magnitude larger than even the most favourable explosion models can provide. In an inhomogeneous model, on the other hand, the strength of the magnetic field can be substantially smaller (see Equation \ref{eq2.8a}). This makes possible a lower thermal energy density behind the forward shock, which then implies a lower kinetic energy of the expanding envelope.

Since observations indicate an inhomogeneous source structure for SN\,1993J, one may consider the implications for a subcomponent, in which the standard assumptions apply, i.e., equipartition between relativistic electrons and magnetic field ($U_{\rm rel} = U_{\rm B}$), no cooling ($R_{||}/R = 1/2$) and $\gamma_{\rm min} = 1$.  This yields a covering factor $\eta_{\rm cov} = 0.46$ together with $y=0.35$ (see Equation \ref{eq2.1}). Furthermore, since $B\propto \eta_{\rm cov}^2$ (Equation \ref{eq2.8a}), the strength of the magnetic field would decrease by a factor 4.8. Such a lowering of the $B$-value is consistent with the conclusions drawn from observations in \cite{bjo15} regarding the physical properties of the radio emission region in SN\,1993J;  for example, that radiative cooling ceases to be important no later than 100\,days.

\section{SN\,2011dh}\label{sect3}
In addition to a spherically symmetric, homogeneous source, the standard model assumes the evolution of  supernovae to be described by the self-similar solution found by \cite{che82a} \citep[and independently by][] {nad85}. Following the notation in \cite{t/m99}, the validity of this solution defines the CN-phase and is applicable at times late enough that the initial conditions no longer affect the dynamics.

SN\,2011dh was detected at an early date in the radio \citep{sod12,kra12,hor13}. At a later time, also VLBI-observations were made \citep{dew16}; in particular, a spatially resolved image could be obtained at $t = 453$\, days. Combining the radii deduced from the early observations, using the standard model, with the radius directly measured by VLBI, \cite{dew16} found that the forward shock suffered almost no deceleration. In the standard model, this indicated a very steep density gradient in the supernova ejecta; for example, with the ejecta density ($\rho$) varying with velocity ($v$) as $\rho \propto v^{\rm -n}$, observations suggested n\,$\approx$\,28. Furthermore, the deduced velocity of the forward shock was $v_{\rm b}\approx 2.0 \times 10^4$\,km/s.

Optical observations were also done at an early stage; in particular, detailed spectra were obtained that overlapped in time with those in the radio \citep{erg14,mar14}. Of special interest here are the velocities that could be deduced from the absorption lines. These lines originate in the un-shocked ejecta and, thus, should give a lower limit to the velocity of the forward shock. From the earliest spectra (3-4 days), \cite{erg14} stated that the H$\alpha$-line showed absorption at least up to $2.5\times 10^4$\,km/s. The reason for the lower limit is the interference by another line, which starts to become important at the frequency corresponding to that velocity. \cite{mar14} modelled this spectral region in some detail and found indications of absorption stretching up to  $3.0\times 10^4$\,km/s (see their fig.\,8).

The absorption line profiles reflect the density gradient in the ejecta. The modelling in \cite{mar14} best reproduced the observations for n=6.  Furthermore, \cite{jer15} found the same value of n to be appropriate for the spectra at later times. Hence, while radio observations indicate a very steep density gradient for the ejecta, optical observations are best reproduced with a shallow slope. Since it is not clear how the absorption line profiles can be accounted for with a steep density gradient, one may consider the consequences for the radio emission region assuming n\,$\approx 6$. 
The shock velocity varies with time as $v_{\rm b} \propto t^{-1/(n-2)}$. With n\,$\approx 6$, extrapolating back from the VLBI-observation at $t = 453$\,days to the first radio observations by \cite{hor13} at $t \approx 4$\,days, one finds a velocity, roughly, three times higher than obtained with the standard model (cf. Figure \ref{fig4}). This, then, indicates an inhomogeneous source with a low covering factor ($\eta_{\rm cov} \approx 0.1$).

An inhomogeneous source structure is also indicated by the flat-topped spectra found by \cite{kra12}. In order to obtain acceptable $\chi^2$-values from their fits to the standard model, they needed to artificially increase their measurement errors by factors 3-7. Even so, they found excess emission below the deduced spectral peak and, furthermore, it was systematically $\approx 10\,\%$ lower than the observed spectral maximum.

The lower limit for the velocity of the forward shock obtained from the absorption lines is an underestimate for two reasons: (1) The maxim ejecta velocity corresponds to that at the reverse shock, i.e., $R_{\rm r}/t$, where $R_{\rm r}$ is the position of the reverse shock. In the standard model, the velocity of the forward shock is $ [(n-3)/(n-2)]R_{\rm b}/t$, where $ R_{\rm b}$ is the position of the forward shock. Hence, their ratio is $ [(n-3)/(n-2)]R_{\rm b}/R_{\rm r}$. It is seen from \cite{che82a} that this value is $\approx 1.1$ and, in particular, it is almost independent of n. As a result, the velocity of the forward shock is expected to be roughly $10\,\%$ larger than the maximum ejecta velocity. (2) In the modelling of \cite{erg14} and \cite{mar14}, the X-ray emission from the reverse shock was not included. As shown by \cite{c/f94}, this emission causes almost complete ionisation of hydrogen in the ejecta closest to the reverse shock. Although the extent of this ionisation is hard to estimate, it is likely that the high velocity wing of the hydrogen absorption corresponds to ejecta some distance away from the reverse shock, i.e., at a velocity lower than that at the reverse shock.

\subsection{The initial conditions and the unified solution}\label{sect3a}
It is clear that extrapolating the standard model backwards in time, there will come a moment when the initial conditions cannot be ignored. \cite{h/s84} showed that the initial phase of the evolution can also be described by a self-similar solution (the piston phase). This was the starting point for \cite{t/m99} to find a unified solution that coupled the initial ejecta dominated phase (i.e., the presence of a reverse shock) to the later Sedov-Taylor phase (i.e., no reverse shock). They argued that the dynamical state of many observed supernova remnants corresponds to this transition. In addition to $\ell\equiv R_{\rm b}/R_{\rm r}$, they introduced a parameter $\phi$, which is the ratio of the pressures behind the reverse and forward shocks. They showed that in the limit of constant values for both of these parameters, analytical solutions could be found that described the whole evolution for different values of n. Furthermore, these analytical solutions were excellent approximations to the numerical results, which they obtained from solving the relevant hydrodynamical equations.

Although the main interest of \cite{t/m99} was to apply their analytical results to supernova remnants, they mentioned  that the self-similar solution of \cite{h/s84} transitions to another self-similar solution within the ejecta dominated phase (i.e., before the Sedov-Taylor phase). This latter solution corresponds to that obtained by \cite{che82a}. Hence, the solution found by \cite{t/m99} explicitly accounts for the evolution of the initial piston phase into a phase where the dynamics is determined by balancing the momentum input at the forward shock to that at the reverse shock \citep{che82b}. 

In the initial piston phase, the velocity of the forward shock is constant and equal to $R_{\rm b}/t$. Hence, the ratio between this velocity and the ejecta velocity at the reverse is given by $\ell$. From their self-similar solution, \cite{h/s84} determined that $\ell = 1.1$ (The more extensive discussion in \cite{t/m99} arrived at the same value). One may note that this value is the same as that appropriate for the later CN-phase (see above). Hence, the ratio of the velocity of the forward shock and the ejecta velocity at the reverse shock remains roughly constant during the whole ejecta dominated phase.

The transition between the two self-similar phases was discussed in more detailed by \cite{bjo25}. It was found that for n $>$ 5 and a constant mass-loss rate from the progenitor star, the transition occurs at a characterised time given by
\begin{equation}
	t_{\rm CN} = \frac{2\,(n-5)}{\phi_{\rm ED}(n-3)^2} \frac{w_{\rm core}^{n-5}}{[n/5-w_{\rm core}^{n-5}]} \frac{E}{v_{\rm ej}^3} \frac{v_{\rm w}}{\dot{M}_{\rm w}}. 
	\label{eq3.1} 
\end{equation}
The structure of the ejecta is assumed to have a constant density core and an envelope in which the density varies as $\rho (v) \propto v^{\rm -n}$ between $v=v_{\rm core}$ and a maximum velocity $v=v_{\rm ej}$. The results for other ejecta structures and mass-loss rates can be found in \cite{bjo25}. Furthermore, $w_{\rm core} \equiv v_{\rm core}/v_{\rm ej}$ and $E$ is the total kinetic energy of the ejecta. $\phi_{\rm ED}$ is the ratio of pressures in the ejecta dominated phase. Its value is discussed in \cite{t/m99} but, in most cases, one may use the value appropriate for the piston phase ($\phi_{\rm ED} = 0.34$). $\dot{M}_{\rm w}$ and $v_{\rm w}$ are the mass-loss rate and wind velocity, respectively, of the progenitor star. 

The evolution of the position of the forward shock with time is then given by
\begin{equation}
	\hat{t}(\hat{R}_{\rm b}) = \hat{R}_{\rm b} \left(1+\hat{R}_{\rm b}^{1/2}\right)^{2/(n-3)}.
	\label{eq3.2}
\end{equation}
Here, $\hat{t} \equiv t/t_{\rm CN}$ and $\hat{R_b} \equiv R_b/R_{b,\rm {CN}}$, where $R_{b,{\rm CN}} = \ell_{\rm ED} v_{\rm ej} t_{\rm CN}$ and $\ell_{\rm ED} = 1.1$ is the value of $\ell$ in the ejecta dominated phase.

With the definitions $v_{\rm b} \equiv R_{\rm b}/t$ and $v_{\rm bo} \equiv v_{\rm b}(t=0) = \ell_{\rm ED} v_{\rm ej}$, one finds from Equation (\ref{eq3.2}) that $v_{\rm b}(t)/v_{\rm bo} = 1/(1+\hat{R}_{\rm b}^{1/2})^{\rm 2/(n-3)}$. The evolution of the forward shock can then be written
\begin{equation}
	\hat{t} = \left[1-\left(\frac{v_{\rm b}}{v_{\rm bo}}\right)^{\rm (n-3)/2}\right]^2 \left(\frac{v_{\rm bo}}{v_{\rm b}}\right)^{\rm (n-2)}.
	\label{eq3.3}
\end{equation}
This function is plotted in Figure \ref{fig3} and shows the transition from the self-similar solution of \cite{h/s84} to that of \cite{che82a}. The limiting values as $t\rightarrow 0$ (HS-solution) and $t\rightarrow \infty$ (CN-solution) are indicated.

 \begin{figure}[h!]
	\includegraphics[width=\linewidth]{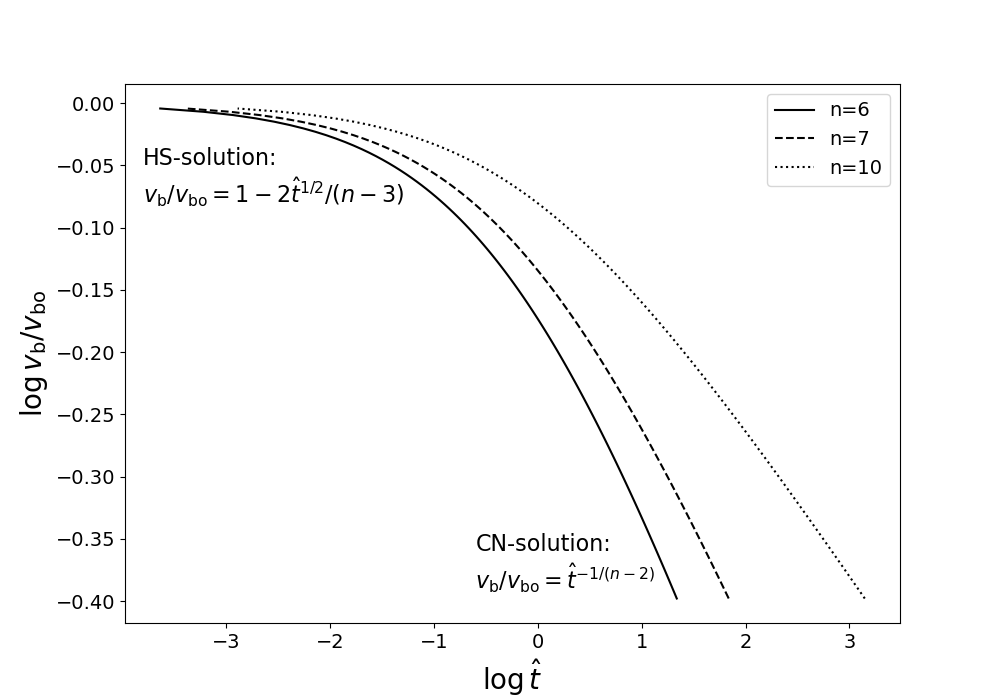}
	\caption{The deceleration of the forward shock with time in the transition between the initial piston phase (HS-solution) and the later phase, which is independent of the initial conditions (CN-solution).}
	\label{fig3}
\end{figure}

In order to compare with observations, $v_{\rm b}$ is shown in Figure \ref{fig4} for different values of $v_{\rm bo}$. The results from \cite{hor13} and \cite{kra12} have all been rescaled to a distance of 7.8\,Mpc to SN\,2011dh \citep{erg14}. Furthermore, the value of $R_{\rm b}$ was deduced from observations assuming the standard model to apply. The curves are all normalised so that they coincide with the VLBI-value measured at $t=453$\,days ($v_{\rm b} = 1.9\times10^4$\,km/s). Note that this normalisation determines $t_{\rm CN}$ for each value of $v_{\rm bo}$. In principle then, the covering factor is obtained from the difference between the curves and observations. However, radio observations alone cannot determine which of the curves is the relevant one. 

\begin{figure}[h]
	\includegraphics[width=\linewidth]{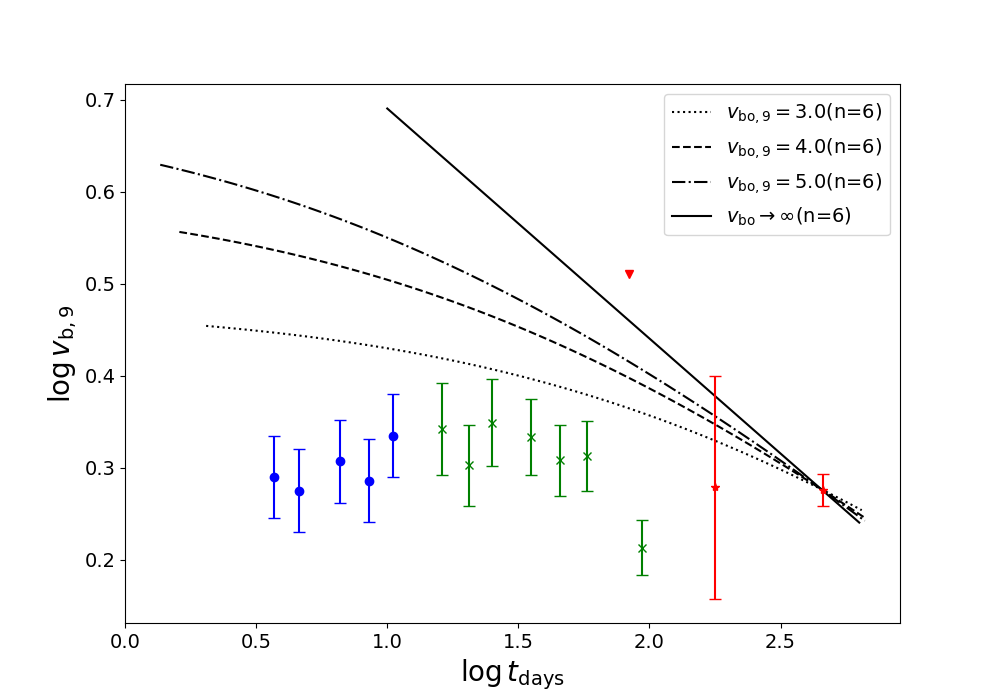}
	\caption{The position of the forward shock for various initial velocities. The curves have all been normalised so that they coincide with the VLBI-measured value on day 453. Also shown are the observed values obtained by \cite{hor13} (blue $\bullet$), \cite{kra12} (green $\times$) and \cite{dew16} (red $\ast$ and an upper limit,  red $\triangledown$).}
	\label{fig4}
\end{figure}

The needed constraints can be provided by the optical observations. As mentioned above, the modelling of \cite{mar14} indicated that H$\alpha$-absorption could be discerned up to $3.0 \times 10^4$\,km/s during, roughly, the first 4-10\,days. With $\ell_{\rm ED} = 1.1$, one finds $v_{\rm b,9}\,(\equiv v_{\rm b}/10^9) = 3.3$ during this period. It is seen from Figure \ref{fig4} that this corresponds to $v_{\rm bo,9}\,(\equiv v_{\rm bo}/10^9) \approx 4$ and a covering factor $\eta_{\rm cov} \approx 0.5$. Again, as mentioned above, this should be taken as a lower limit to $v_{\rm bo,9}$ and an upper limit to $\eta_{\rm cov}$, since the effects of the X-ray ionisation were not taken into account. Furthermore, one finds from Equation (\ref{eq3.3}) that $t_{\rm CN} = 50$\,days for $v_{\rm bo,9} = 4.0$ and $t_{\rm CN} = 16$\,days for $v_{\rm bo,9} = 5.0$.

It is seen from Figure \ref{fig4} that the values of $v_{\rm b}$ deduced by \cite{hor13} for $t\,\lsim\,10$\,days are somewhat smaller as compared to those obtained by \cite{kra12}  during their first observations for $t\,\gsim\,16$\,days. This could be due to a combination of an inhomogeneous source structure and the fitting procedure. As already mentioned, the spectra presented by \cite{kra12} are flat-topped, indicating the overlap of several subcomponents. When observations are then analysed with the use of a homogeneous source model, fitting criteria determine which of the subcomponents is selected. \cite{kra12} mention that their fitting procedure systematically gives a spectral peak, which is roughly 10\,\% smaller than the observed spectral maximum. \cite{hor13} give no details about their fitting criteria and, in addition, they deduce the source parameters using light curves. Hence, the inhomogeneous source structure could cause these different fitting procedures to result in somewhat different source parameters. For example, if the peak of a light curve in \cite{hor13} corresponded to the observed spectral maximum as observed by \cite{kra12}, their deduced value of $\nu_{\rm abs}$ would be systematically $\approx 10\,\%$ higher as compared to the one obtained from the fitting criteria used by \cite{kra12}. Hence, the deduced value for $R$ would be $\approx 10\,\%$ lower (see Equation \ref{eq2.4}). If the deduced $R-$values are renormalised by this amount, one finds that the values of $v_{\rm b}$ in Figure \ref{fig4} is roughly constant during the first 20-30\,days. This is similar to the evolution of SN\, 1993J during the first few hundred days.

\subsection{The brightness temperature of SN\,2011dh}\label{sect3b}
Late time spectral fitting of SN\,2011dh has been performed by \cite{ven23} (497\,days and 1197\,days). The first of these is rather close to the VLBI-observations of \cite{dew16} done on day 453. Hence, it is possible to deduce a rough estimate of the brightness temperature for this period of the supernova evolution.

The deduced peak spectral flux is roughly the same for the two occasions. Therefore, it will be assumed that the peak spectral flux is also the same on day 453. Converting the value given in \cite{ven23} to the one corresponding to the spectral peak gives $f_{\nu_{\rm abs}} = 4.7$\,mJy. Assuming the self-absorption frequency to scale as $\nu_{\rm abs} \propto R^{-1}$, yields $\nu_{\rm abs,10} = 0.079$ on day 453. With the value measured by \cite{dew16}, $\theta = 0.636$\,mas, one then finds from Equation (\ref{eq2.5}) $T_{\nu,10}^{\rm br,obs} = 0.82/\eta_{\rm cov}$. 
The optically thin emission at these late times is well fitted by p=2.7, which is the same as for SN\,1993J. With a distance of 7.8\,Mpc, the observed peak spectral flux corresponds to $F_{\nu_{\rm abs,27}} = 0.34$. The brightness temperature expected from a synchrotron source is obtained from Equation (\ref{eq2.6}) as $T_{\nu,10}^{\rm br,synch} = 3.0\,y^{1/9.2}/\eta_{\rm cov}^{1/18.4}$. Again, equating these two brightness temperatures results in $y\eta^{8.7} = 5.9 \times 10^{-6}$. Assuming the same value for $y$ in SN\,2011dh as in SN\,1993J, the covering factor is a factor 1.6 smaller in the former as compared to the latter. It is interesting to note that the peak spectral flux has decreased in SN\,2011dh at these late times as compared to the roughly constant value during the first $\approx 100$\,days. The average value for the initial peak spectral flux is $F_{\nu_{\rm abs,27}} = 0.54$, which is a factor 1.6 larger than  the value at day 453\,days. Hence, it is possible that the decreasing spectral flux in SN\,2011dh is due to a covering factor, which decreases by a factor 1.6 between $\sim 100$ and $\sim 400$\, days. If so, it implies that the covering factor in SN\,2011dh during the first $\approx 100$\,days was roughly the same as that in SN\,1993J.

\subsection{The evolution of the synchrotron self-absorption frequency}\label{sect3c}
In a homogeneous source, the evolution of $F_{\nu_{\rm abs}}$ and $\nu_{\rm abs}$ are related, since they both reflect the varying source properties. This is not so for an inhomogeneous source; the value of $\nu_{\rm abs}$ is determined by the conditions along the line of sight in the relevant subcomponent, while $F_{\nu_{\rm abs}}$ depends also on its covering factor. As discussed above, when  $F_{\nu_{\rm abs}}$ is constant and $\nu_{\rm abs}\propto 1/R$, observations yield a constant value also for $y\eta_{\rm cov}^{\rm p+6}$. One may note that for a homogeneous source, it can usually be argued that a constant value for $F_{\nu_{\rm abs}}$ implies $\nu_{\rm abs} \propto 1/R$, due to its insensitivity to the value of $y$ (see Equation \ref{eq2.4}). This is not possible for an inhomogeneous source, since unknown variations in the covering factor would prevent such an inference. This shows that by comparing the evolutions of $F_{\nu_{\rm abs}}$ and $\nu_{\rm abs}$, information can be obtained regarding the variation of the covering factor.

The observed flat-topped spectra in SN\,2011dh clearly indicated an inhomogeneous source structure. Hence, the use of a homogeneous source model to fit the data implies that the value of $F_{\nu_{\rm abs}}$ and, in particular, the value of $\nu_{\rm abs}$ are sensitive to the fitting procedure. The values obtained in \cite{ven23} result from anchoring the fit to the well-observed optically thin part of the spectrum. These values, then, apply to the subcomponent, which dominates the optically thin emission. On the other hand, the fitting criteria chosen by \cite{kra12} (see Section \ref{sect3a}) are likely to result in values, corresponding to the average ones in the distribution of subcomponents. As already mentioned, they noted that the observed spectral maximum was systematically $\approx 10\,\%$ larger than the value they deduced for $\nu_{\rm abs}$. In an attempt for a more realistic evaluation of the evolution of $\nu_{\rm abs}$, it will be assumed that their observed spectral maximum corresponds to the component dominating the optically thin emission. Therefore, in the following, the values found for $\nu_{\rm abs}$ in \cite{kra12} will be multiplied by 1.1. Note also that the values for $F_{\nu_{\rm abs}}$ and $\nu_{\rm abs}$ correspond to the spectral peak, while those given in \cite{kra12} are measured at $\tau=1$. The conversion factors between the two ways of characterising the spectrum are discussed in \cite{kra12}.

\begin{figure}[h!]
	\includegraphics[width=\linewidth]{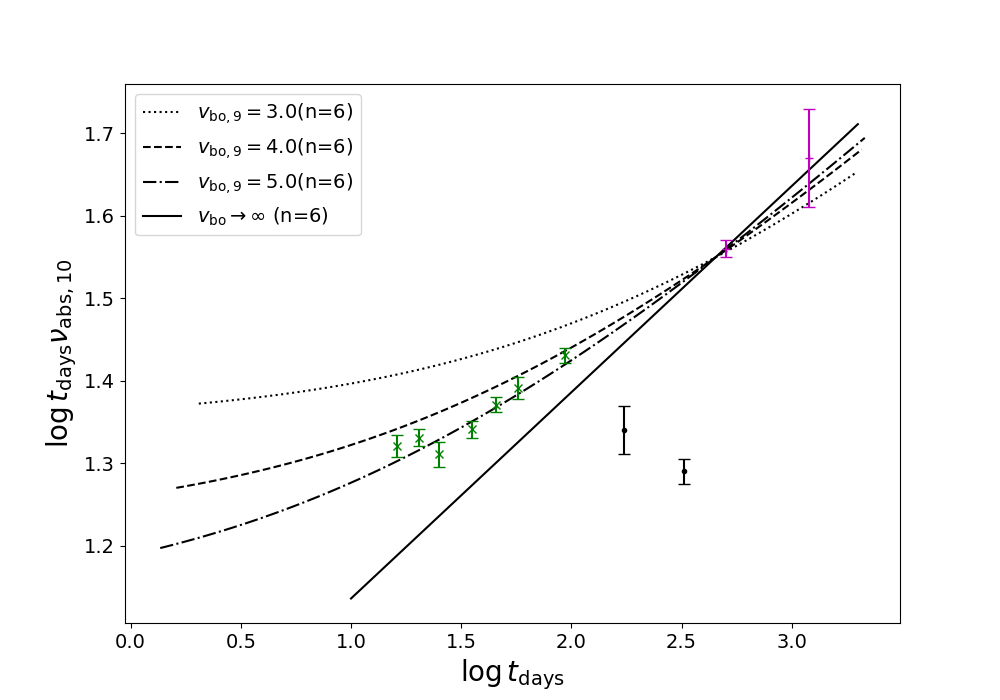}
	\caption{The evolution of the self-absorption frequency ($\nu_{\rm abs}$) for various initial velocities. The curves have all been normalised so that they coincide with the deduced  value on day 453 (see text). Also shown are the observed values by \cite{kra12} (green $\times$) and \cite{ven23} (magenta $+$). The observed peak of the light curves at $\nu = 1.3$\,GHz and $\nu = 0.61$\,GHz are from \cite{yad16} (black $\cdot$). As discussed in the text, their lower values are due to the decreasing  peak spectral flux, which causes the flux at a given frequency to peak before that frequency corresponds to the self-absorption frequency.}
	\label{fig5}
\end{figure}

The evolution of $\nu_{\rm abs}$ can be obtained from Equation (\ref{eq2.7}) as
\begin{equation}
	\nu_{\rm abs,10}t_{\rm day} = 1.2 \times 10\,\frac{\hat{R}(\rm p) F_{\nu_{\rm abs},27}^{\rm (p+6)/(2p+13)}}{\left(y\eta_{\rm cov}^{\rm p+6}\right)^{\rm 1/(2p+13)}}
	\frac{\left(1+\hat{R}_{\rm b}^{\rm 1/2}\right)^{\rm 2/(n-3)}}{v_{\rm bo,9}}.
	\label{eq3.4}
\end{equation}
With $\hat{R}(2.7) = 5.0$ from Figure \ref{fig2}b and n = 6, the result is shown in Figure \ref{fig5} for various values of $v_{\rm bo,9}$ together with the observed values. No values for $\nu_{\rm abs}$ are given by \cite{hor13}. Since they do not provide details of their fitting procedure, no attempt has been made to deduce values for this initial period. Here,  $F_{\nu_{\rm abs},27} = 0.34$ and $y\eta^{8.7} = 5.9 \times 10^{-6}$ have been used (see Section \ref{sect3b}). One may note that $\nu_{\rm abs} t \propto 1/v_{\rm b}$, so that the curves in Figure \ref{fig5} are basically the inverse of those in Figure \ref{fig4} and, furthermore, that the normalisation is such that their values of $\nu_{\rm abs}$ at $t = 453$\,days are the same and correspond to that deduced from observations (i.e., the same normalisation as in Figure \ref{fig4}).

The important thing to note from Figure \ref{fig5} is that the overall evolution of $\nu_{\rm abs}$ is rather well described by these curves. This implies  that $F_{\nu_{\rm abs}}/(y^{\rm 1/(p+6)}\eta_{\rm cov})$ is roughly constant during the whole observing period, which leads to $\nu_{\rm abs} \propto 1/R$; in particular, the evolution of $\nu_{\rm abs}$ does not seem to be affected by the decreasing value of  $F_{\nu_{\rm abs}}$, which suggests that this decrease is due mainly to a decreasing covering factor. This agrees with the conclusion in Section \ref{sect3b}. It should be noted, though, that these results are independent of each other; the one in  Section \ref{sect3b} is based on a comparison of the brightness temperature at one particular time ($t= 453$\,days) to that in SN\,1993J, while the one here is deduced from the evolution of $\nu_{\rm abs}$ in SN\,2011dh during the whole observing period.

Due to the insensitivity to $y$, it is likely that $\eta_{\rm cov}$ is constant during the first $\approx 100$\,days when $F_{\nu_{\rm abs}}$ is constant, while it decreases by a factor 1.6 thereafter. However, it is interesting to consider the structural changes needed for the decrease in $F_{\nu_{\rm abs}}$ to be consistent with a constant value for $\eta_{\rm cov}$. With $y \propto F_{\nu_{\rm abs}}^{\rm p+6}$ and p = 2.7, this yields a value of $y$ smaller by a factor 60. Furthermore, since $\gamma_{\rm abs} \propto F_{\nu_{\rm abs}}$, this leads to $B \propto F_{\nu_{\rm abs}}^{-2}$ and $U_{\rm rel} \propto F_{\nu_{\rm abs}}^{\rm p+4}$. This implies a value for $B$ a factor 2.6 larger and the value of $U_{\rm rel}$ to be smaller by a factor of 23. Although such a substantial change to the source structure cannot be excluded, the most straightforward explanation for the decreasing value of $F_{\nu_{\rm abs}}$ is that it is due mainly to a decreasing covering  factor.

The value of $v_{\rm bo,9}$ can be constrained from Figure \ref{fig5} to lie in the range 4.0 - 5.0. This is somewhat higher than the velocity deduced directly from the optical observations. As discussed in Section \ref{sect3a}, this indicates that X-ray emission ionised the hydrogen close to the reverse shock.  The value for $y\eta^{8.7}$ during the first $\approx 100$\,days in SN\,2011dh is roughly  the same as that in SN\,1993J. The use of the standard assumptions regarding the structure of the dominating subcomponent (as discussed in Section \ref{sect2c}), results in $\eta_{\rm cov} = 0.45$, which decreases by a factor 1.6 to $\eta_{\rm cov} = 0.28$ during the later stages in the evolution of SN\,2011dh.

\subsection{The optically thin light curve of SN\,2011dh}\label{sect3d}
The optically thin light curve for $\nu=8.4$\,GHz  has a concave break at a few hundred days. \cite{dew16} have suggested that this is due to the appearance of a new component. They also point out that the initial decline of the light curve is unusually steep. One may note that this break occurs during the period when the peak spectral flux declined. It is therefore possible, that not only the late time flattening of the light curve but also its initial steep decline may instead be caused by the transition between two phases characterised by different but constant values of $F_{\nu_{\rm abs}}$.

The optically thin light curve can then be written
\begin{equation}
	F_{\nu}(t) = F_{\nu_{\rm abs}}^{\rm thin} \left(\frac{\nu_{\rm abs}}{\nu}\right)^{\rm (p-1)/2)} \Lambda (t),
	\label{eq3.5}
\end{equation}
where $F_{\nu_{\rm abs}}^{\rm thin} = F_{\nu_{\rm abs}} \tau_{\rm abs} (\rm p)/[1-\tau_{\rm abs}(\rm p)]$ is the optically thin spectral flux extrapolated to $\nu_{\rm abs}$. Furthermore, $\Lambda (t)$ describes the transition between the two phases. It will be parameterised as
\begin{equation}
	\Lambda (t) = \frac{\left\{\exp[-(t/t_{o})^{\rm m}] + \delta\right\}}{1+\delta}.
	\label{eq3.6}
\end{equation}
Here, $\delta/(1+\delta)$ is the ratio between the constant values of $F_{\nu_{\rm abs}}$ in the two phases, $t_{o}$ is the time for the transition and m measures its steepness. 

 \begin{figure}[h!]
	\includegraphics[width=\linewidth]{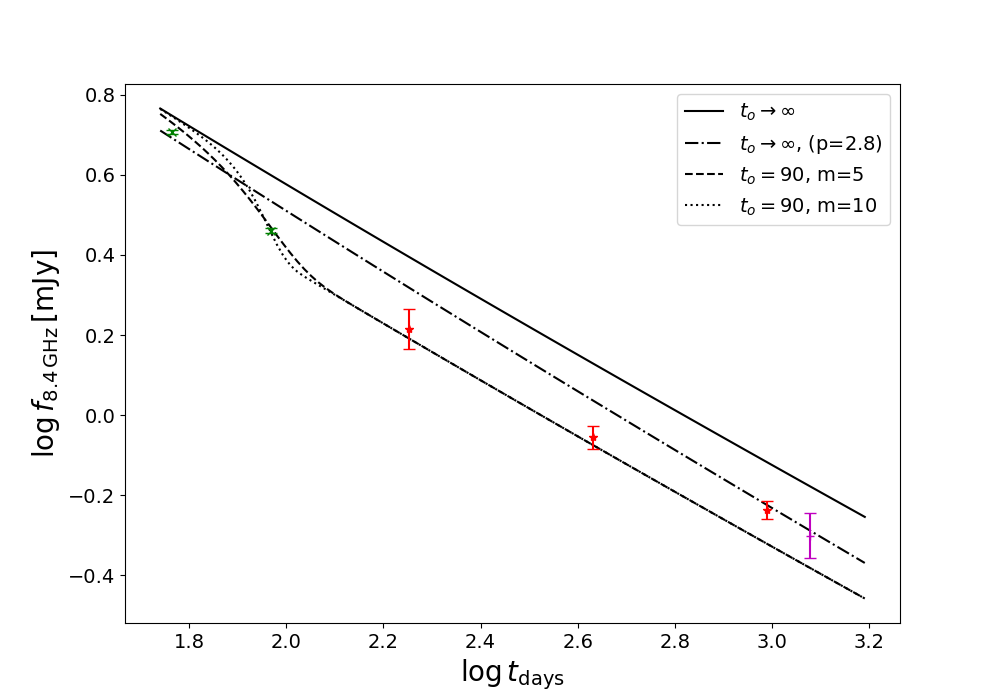}
	\caption{The optically thin light curve for $\nu = 8.4$\,GHz and p = 2.7. The solid line corresponds to a constant peak spectral flux (i.e., $t_{o} \rightarrow \infty$). Although the rather rapid decline initially constrains the value of $t_{o}$, the duration of the transition is less well determined (i.e., the value of m). Also shown is the light curve for p = 2.8 and a constant peak spectral flux. The observations are from \cite{kra12} (green $\times$), \cite{dew16} (red $\ast$) and \cite{ven23} (magenta +).}
	\label{fig6}
\end{figure}

With the values $F_{\nu_{\rm abs}}$ and $y\eta^{8.7}$ derived above, one finds from Equation (\ref{eq2.4}) that $\nu_{{\rm abs},10} = 5.8/R_{15}$. Note that this  value corresponds to the subcomponent, which dominates the optically thin emission (see Section \ref{sect3c}). Furthermore,  the observed drop in spectral flux by a factor 1.6 yields $\delta = 1.7$. With $\nu = 8.4$\,GHz, Equation (\ref{eq3.5}) becomes
\begin{equation}
	F_{\rm 8.4GHz}(t) = 2.5 \frac{F_{\nu_{\rm abs}}}{R_{15}^{0.85}(t)}\left\{\exp[-(t/t_{o})^{\rm m}] + 1.7\right\},
	\label{eq3.7}
\end{equation}
where $F_{\nu_{\rm abs}}$ is the peak spectral flux in the first phase.

In order to compare directly with the result in \cite{dew16}, Equation (\ref{eq3.7}) is shown in Figure \ref{fig6} using the observed spectral flux $f_{\nu}$ instead of $F_{\nu}$; i.e.,  $F_{\nu_{\rm abs}}  \rightarrow f_{\nu_{\rm abs}} = 7.5$\,mJy, which is the average of the peak spectral flux values as given by \cite{kra12}. Since the characteristics of the transition between the two phases are determined mainly by $t_{o}$ and m, $v_{\rm bo} = 4.0$ and n=6 have been used.

The curves with $t_{o} \rightarrow \infty$ correspond to $\Lambda (t) =1$ and can be used as reference for the effects due to a declining spectral peak flux; i.e., the difference between a reference curve and the other curves gives the value of $\Lambda (t)$ at a given time. It is seen that the most rapid drop in $f_{\rm 8.4\,GHz}(t)$ occurs around day 93, which indicates $t_{o} \approx 90$\,days. Although optical depth effects are unlikely to be important at this time, they may lower the optically thin spectral flux at the first observation on day 58, i.e., the absorption corrected spectral flux would be higher than the observed value shown in Figure \ref{fig6}. Another reason for the observed spectral flux to  lie below the reference curve is a value of m sufficiently small that $\Lambda(t) < 1$ already at this early date.

 As mentioned in Section \ref{sect3a}, the inhomogeneous source structure causes the fitting procedure used in \cite{kra12} to give a value for $\nu_{\rm abs}$, which is $\approx 10\,\%$ lower than the observed peak spectral flux. On day 58, the latter is then $\approx 4.2$\,GHz, which shows that the light curve frequency is only a factor $\approx 2$ larger than the observed peak frequency. This gives a non-negligible contribution to the absorption. The value of m can then be determined by requiring  the light curve to go through the absorption corrected flux; this leads to m\,$\approx 5$. This is a lower limit to m (for $t_{o} = 90$\,days), since additional contributions to the absorption from components with peak frequencies above $\approx 4.2$\,GHz cannot be excluded; hence, m\,$\gsim\,5$. All the curves in Figure \ref{fig6} are drawn for p\,=\, 2.7, except for one, which is the reference curve for p\,=\,2.8. This shows that in order to account for the spectral flux on day 58, p\,$\lsim$\,2.8 is indicated.
 
 It is seen in Figure \ref{fig6} that the simple parameterisation used for the transition underestimates the flux at late times  ($\gsim\,1000$\,days). This suggests that the transition is more complex than described by $\Lambda (t)$ in Equation (\ref{eq3.6}). A possible cause for these deviations is that the peak spectral flux in the late phase is not constant, e.g., due to  a varying covering factor. 
 
Although no spectra are available for the time period when the peak spectral flux decreased, \cite{yad16} obtained light curves for 1.3\,GHz and 0.61\,GHz, which peaked at $t = 174$\,days and $t = 323$\,days, respectively. The corresponding spectral fluxes at the peak were $6.20 \pm 0.16$\,mJy and $4.51\pm0.23$\,mJy. As discussed above, when $F_{\nu_{\rm abs}}$ is constant and $\nu_{\rm abs} \propto 1/R$, the peak of the light curve for a given frequency $\nu$ occurs at the same time as this frequency corresponds to the spectral peak. This is not so for a varying value of  $F_{\nu_{\rm abs}}$. As an example, when $F_{\nu_{\rm abs}}$ decreases with time, the light curve for a given frequency peaks before that frequency corresponds to the spectral peak; a faster decrease of $F_{\nu_{\rm abs}}$ results in a longer time interval between the occurrences of the two peaks. Furthermore, inhomogeneities are likely to increase this time interval; for a spectrum like that of SN\,2011dh, the light curve peak is then due to the subcomponent with the lowest value of $\nu_{\rm abs}$, while the spectral peak corresponds to a subcomponent with a higher value for  $\nu_{\rm abs}$. This difference can be significant, as is shown in Figure \ref{fig5}, where the results from \cite{yad16} are included.

\section{Discussion}\label{sect4}
The structural properties of the supernova envelope can be probed either by optical or radio observations; in particular, this is so when the observations are done early on. In the optical, absorption line profiles then give information of both velocity and density of the outermost part of the ejecta, while radio observations can be used to derive the velocity of the forward shock. Hence, a comparison of the deduced properties constrains the model, which connects the ejecta properties to those of the forward shock. It is usually assumed that the radio source is homogeneous and that the dynamics of the shocked region is governed by the self-similar solution appropriate for a late phase when the initial conditions no longer effect the evolution \cite{che82a}. 

It is quite rare that observations of sufficient quality are done quasi-simultaneously in the two wavelength regimes to allow such a comparison. Here, SN\,2011dh stands out, since extensive observations were done during the first 100\,days both in the optical \citep{erg14,mar14} and radio \citep{sod12,hor13,kra12}. Furthermore, the radio source was spatially resolved at a later stage \citep[$t\,=\,453$\,days,][]{dew16}. It is noteworthy that the deduced ejecta structures differ dramatically. The optical observations indicated a shallow density structure in the ejecta, while the radio observations pointed to a very steep one. In addition, the optical observations implied a substantially higher velocity for the forward shock than obtained from the radio observations. 

In order to overcome these differences, it was shown in Section \ref{sect3} that two changes need to be done in the model normally used to interpret the radio observations. A shallow density structure is consistent with the radio observations, if the initial interaction between supernova ejecta and circumstellar medium corresponds to the piston phase (i.e., constant shock velocity). The transition to the traditional model occurs at $\sim 50$\,days. The low velocity of the forward shock deduced from the radio observations is due to the assumption of a homogeneous source structure. Consistency with the optical observations is obtained with an initial covering factor for the radio source of $\eta_{\rm cov} \approx 0.5$.

As is shown in Section \ref{sect3b}, assuming a homogeneous source when, in fact, the source is inhomogeneous will substantially affect the deduced values of the source parameters; for example, the value of the magnetic field varies as $B\propto \eta_{\rm cov}^2$. Hence, assuming a homogeneous source will overestimate the strength of the magnetic field, which, in turn, leads to an underestimate of the energy density of the relativistic electrons. One may note that this is opposite to the conclusion reached in a situation, where the inverse Compton scattered X-ray emission is observed instead of the spatially resolved source radius; i.e., neglecting the effects from an inhomogeneous source structure will then cause an overestimate of the ratio between the energy densities in relativistic electrons and magnetic field \citep{bjo24}. The reason for this difference is that in the first case, a covering factor less than unity increases the brightness temperature, which, in turn, lowers the strength of the magnetic field in the subcomponent. On the other hand, in the latter situation, the inhomogeneities imply that there are electrons, which may contribute to X-ray but not the radio emission.

It was found in Section \ref{sect3c} that the self-absorption frequency varied inversely with the deduced radius during the whole observing period of SN\,2011dh. One should note that this inference relies on the presence of the transition between the initial piston phase and the later phase relevant for the standard model. This gives the curvature needed for the time evolution of the self-absorption frequency to be consistent with $\nu_{\rm abs} \propto 1/R$ (see Figure \ref{fig5}). Together with the observed constant value of the peak spectral flux during the first $\approx 100$\,days, this implies $B \propto 1/R$ during this time period. This is in contrast to the often used assumption that a constant fraction of the thermal energy density behind the forward shock goes into amplifying the magnetic field, which yields instead $B \propto 1/t$. These two scaling relations imply rather different mechanisms for the amplification of the magnetic field.

One may note that in the piston phase ($R\,\propsim\,t$) or in the late phase with a steep ejecta density structure ($R\,\propsim\,t^{(n-3)/(n-2)}$), it is hard to distinguish between these two scaling relations. However, the shallow density slope deduced for SN\,2011dh (n $\approx 6$) offers an opportunity to compare the expected evolution in the two cases. For $B\propto 1/t$ and $\nu_{\rm abs}\propto 1/R$, one finds that $\gamma_{\rm abs} \propto (t/R)^{1/2}$. Since $F_{\nu_{\rm abs}} \propto  (\nu_{\rm abs}R)^2 \gamma_{\rm abs}$, this leads to $F_{\nu_{\rm abs}} \propto  (t/R)^{1/2}$. Hence, with $B\propto 1/t$, an increase in the peak spectral flux is expected throughout the whole observed period. This is contrary, not only to the initial phase with a constant value, but, in particular, to the later decline. Instead, with $B\propto1/R$, $\gamma_{\rm abs}$ is constant and the later decline can be attributed to a decreasing covering factor.

It is often observed that supernovae have extended periods with $F_{\nu_{\rm abs}}(t)\approx$ constant. One may note that a similar characteristic is associated with the jets of compact extragalactic radio sources. Here, the observed flat radio spectra are found to be due to overlapping subcomponents with different $\nu_{\rm abs}$ but the same value of $F_{\nu_{\rm abs}}$. This has been called a cosmic conspiracy by \cite{cot80}. The first models of the radio emission from extragalactic radio sources assumed a steady adiabatic outflow of relativistic electrons and magnetic fields from the region around the central black hole \citep{van66}. It was soon realised that such models predicted a radial decline in the radio emission, which was too rapid to be consistent with observations. This was due to adiabatic cooling. \cite{b/k79} suggested that re-acceleration occurred in the jet of the outflowing relativistic electrons, which compensated for the adiabatic losses. They showed that the observations could be accounted for with a model, in which the jet had a constant opening angle so that the emitting surface scaled with radius as $R^2$. Furthermore, since the strength of the radial component of the injected magnetic field is expected to vary as $R^{-2}$, while the perpendicular component should decline slower as $R^{-1}$, the latter would dominate at some distance from the injection site. The re-acceleration of the relativistic electrons was assumed to be such that $\gamma_{\rm min}$ stayed constant (see Section \ref{sect2}). Although the physical mechanism responsible for the re-acceleration is not know in detail, the model is still thought to be relevant.

The main difference is then that in supernovae, the evolution with time of one component is observed, while in extragalactic jets, multiple versions of one component are seen at a given time. Also, the constant injection of relativistic electrons in the jet model corresponds to a constant mass-loss rate of the progenitor star for the supernova. One may note, however, that the less-well understood physics differ in the two scenarios. While $B\propto 1/R$ is a direct consequence of a frozen-in magnetic field in the jet model, it is the main unknown for the supernova model. Likewise, while a constant value for $\gamma_{\rm min}$ can be reasonably  well motivated by Fermi-acceleration at the forward shock in a supernova, it is not so clear how re-acceleration can rather precisely balance adiabatic losses in the jet.

The optically thin light curve for 8.4\,GHz began to decline quite steeply at $\approx 90$\,days after which a flattening started after a few hundred days. \cite{dew16} suggested that this late time flattening was due to the appearance of a new component. However, as shown in Section \ref{sect3d}, this concave break may also result from the decrease in $F_{\nu_{\rm abs}}$, which started at the same time. It was argued that the decline by a factor 1.6 was caused mainly by a decline of the covering factor. The covering factor deduced from observations on day 453, could then be used to obtain a value for the covering factor appropriate for the initial phase when $F_{\nu_{\rm abs}}$ was constant. It was found that this covering factor in SN\,2011dh was basically the same as that deduced for SN\,1993J.

It has been argued above that the discrepancies, regarding the properties of the envelope structure drawn from optical and radio observations of SN\,2011dh, are due to the neglect of the initial piston phase and the effects it has on the evolution of the forward shock. The same conclusion was reached in \cite{bjo25} for SN\,1993J. In addition, both of these well-observed supernovae have several other properties in common; for example, (1) a shallow density structure of the envelope (n\,$\approx 6$), (2)  an inhomogeneous source structure characterised by a similar covering factor (at least initially) and (3) a magnetic field strength scaling as $B\propto 1/R$. In SN\,1993J, the latter relation was obtained directly from observations, while in SN\,2011dh optical observations were used to constrain the evolution of the forward shock.

The main difference is that the maximum ejecta velocity is roughly a factor of two larger in SN\,20011dh as compared to SN\,1993J. \cite{des18} have shown that smaller envelope masses are less extended but reach larger ejecta velocities. Hence, this difference is likely due to the lower envelope mass deduced for SN\,2011dh by \cite{ber12} as compare to SN\,1993J \citep{woo94}. 

\cite{c/s10} suggested that there is a limit to the envelope mass, below which its radius is drastically reduced and that SN\,2011dh belongs to such a compact class of type IIb supernovae. This class would then constitute the transition between type IIb  and stripped envelope supernovae. However, it has been shown that the progenitor of SN\,20011dh was a yellow supergiant \citep{van13} and, thus, had an extended envelope. Since the maximum ejecta velocity in SN\,2011dh [$v_{\rm ej} \approx (4-5) \times 10^4$\,km/s] is close to the velocities deduced for type Ib/Ic supernovae, it is not clear whether a class of compact IIb supernovae exists.

The extensive observations of SN\,1993J and SN\,2011dh allow conclusions to be drawn, which are more far-reaching than for most other radio supernovae. It has been argued that their observed optical characteristics result from explosions in binary systems - SN\,1993J \citep{pod93,woo94} and SN\,2011dh \citep{ben13}. Both are type IIb supernovae and it is often assumed that all such supernovae have a similar origin \cite[see][for a discussion]{van23}. The properties of the circumstellar medium in type IIb supernovae may then differ from those pertaining to other supernova explosions. Since the radio emission is due to the interaction between the ejecta and the circumstellar medium, one might expect type IIb supernovae to be distinct also in the radio regime. However, this does not seem to be the case: (1) The properties of inhomogeneities in other types of supernovae \citep{b/k17,bjo24} do not stand out as qualitatively different from those found here. (2) As discussed in \cite{bjo13}, the observed scaling relation between the X-ray and radio luminosities in type Ib/c supernovae could be due to a smaller covering factor at late times. This is akin to the decreasing covering factor argued to be the cause for the observed light curves in SN\,2011dh. (3) The strength of the magnetic field in SN\,1993J and SN\,2011dh scales with radius (rather than time) as $B\propto 1/R$. This deduction relies, in part, on the constant peak spectral flux. As already mentioned,  extended periods of a constant peak spectral flux are common in other radio supernovae, indicating that the same scaling relation applies also for these supernovae.

This suggests, in particular, that the importance of inhomogeneities deduced from the observations of SN\,1993J and SN\,2011dh may be generally applicable to all radio supernovae. Hence, there are two aspects to keep in mind when modelling less-well observed radio supernovae: (1) The deduced properties of the emission region are very sensitive to inhomogeneities. (2) Even if the radio spectrum can be well modelled as a homogeneous source, this does not necessarily imply that the deduced radius corresponds to that of the supernova shock, i.e., the inhomogeneities are dominated by one subcomponent. A good example of such a situation is SN\,1993J.

\section{Conclusions}\label{sect5}
The main goal of the present paper was to find a model for the early evolution of SN\,2011dh that could account for the radio as well as the optical observations. It was found that two main changes need to be done to the standard model, namely:

1) The initial evolution corresponded to a piston phase with a transition to the standard model occurring roughly at 50\,days.

2) The synchrotron emitting region was inhomogeneous with a covering factor around 50\%.

\noindent Furthermore, it was shown that these two additions implied the following:

3) The deduced source parameters are very sensitive to inhomogeneities; e.g, a covering factor of 50\% would increase the ratio of the energy densities of relativistic electrons and magnetic field by roughly a factor 500 ($1/\eta_{\rm cov}^{\rm p+6}$).

4) The strong deceleration of the forward shock (n=6) can be used to argue that the strength of the magnetic field scales with radius as $B\propto 1/R$ rather than with time as $B\propto 1/t$. This has implications for the mechanism responsible for the amplification of the magnetic field.

5) The covering factor started to decrease after about 100\,days. It was suggested that this caused the concave break in the radio light curve rather than the occurrence of a new component.

 6) The properties of SN\,2011dh are very similar to those of SN\,1993J, for example, $B\propto 1/R$. The main difference seems to be the smaller envelope mass of SN\,2011dh as compared to SN\,1993J.

\newpage

\appendix

\begin{center}
{\bf Appendix}
\end{center}

\section{A closure relation for a homogenous, spherically symmetric synchrotron source}

The derivation of the closure relation in this Appendix follows the formulation of the basic synchrotron theory in \cite{bjo21}. The synchrotron spectral emissivity per unity volume and solid angle is
\begin{equation}
	j^{\rm s}_{\nu,{\Omega}} = \frac{1}{64 \pi^{3/2}}\frac{\sigma_{\rm T}}{e} ({\rm p-2})\gamma_{\rm min}^{\rm p-2} \chi({\rm p}) U_{\rm rel} B^{\rm  
	(p+1)/2}\left(\frac{\nu_{o}}{\nu}\right)^{\rm (p-1)/2},
	\label{eqA1}
\end{equation}
where $\sigma_{\rm T}$ is the Thomson cross-section and $\nu_{o} \equiv e/(2\pi mc)$. Furthermore,
\begin{equation}
	\chi({\rm p}) = \frac{3^{\rm (p+2)/2}}{\rm p+1}\frac{\Gamma\left(\frac{\rm 3p+19}{12}\right)\Gamma\left(\frac{\rm 3p-1}{12}\right)
	\Gamma\left(\frac{\rm p+5}{4}\right)}{\Gamma\left(\frac{\rm p+7}{4}\right)},
	\label{eqA2}
\end{equation}
where $\Gamma(z)$ is the gamma function.

 The corresponding absorption coefficient is given by
 \begin{equation}
 \mu^{\rm s}_{\nu} = \frac{\pi^{3/2}}{4}\zeta({\rm p})\frac{e}{mc^2}({\rm p-2})\gamma_{\rm min}^{\rm p-2}U_{\rm rel}B^{(p+2)/	2}
 	\left(\frac{\nu_{\rm o}}{\nu}\right)^{\rm (p+4)/2},
	\label{eqA3}
\end{equation}
where
\begin{equation}
	\zeta({\rm p}) = 3^{\rm (p+1)/2}\frac{\Gamma\left(\frac{\rm 3p+22}{12}\right)\Gamma\left(\frac{\rm 3p+2}{12}\right)
	\Gamma\left(\frac{\rm p+6}{4}\right)}{\Gamma\left(\frac{\rm p+8}{4}\right)}.
	\label{eqA4}
\end{equation}
The optical depth at the spectral peak is
\begin{equation}
	\tau_{\rm abs}(\rm p) =  \frac{\sqrt{\pi}}{2^5}\zeta({\rm p}) \frac{e}{mc^2} \frac{yRB}{ \gamma_{\rm abs}^{\rm p+4}},
	\label{eqA5}
\end{equation}
where the function $\tau_{\rm abs}(\rm p)$ can be obtained from
\begin{equation}
	\exp(\tau_{\rm abs}) = 1+\frac{(\rm p+4)\tau_{\rm abs}}{5}.
	\label{eqA6}
\end{equation}

The expression for $\gamma_{\rm abs}$ is then given by
\begin{equation}
	\gamma_{\rm abs} =  \left[\frac{3\pi}{2^{11}}\frac{\zeta^3({\rm p})}{\chi({\rm p})\tau_{\rm abs}^2(\rm p)\{1-\exp(-\tau_{\rm abs})\}}\right]^{\rm 1/
	(2p+13)}y^{2/(\rm 2p+13)}\left(\frac{F_{\nu_{\rm abs}}}{mc^2}\right)^{\rm 1/(2p+13)}.
	 \label{eqA7}
\end{equation}
It is seen from Equation (\ref{eqA6}) that $1-\exp(-\tau_{\rm abs}) = (\rm p+4)\tau_{\rm abs}\exp(-\tau_{\rm abs})/5)$, which leads to
\begin{equation}
	\hat{\gamma}_{\rm abs} (\rm p) = \left\{2.8\times 10^{31}\frac{\zeta^3(\rm p)}{(\rm p+4)\chi(\rm p)\tau_{\rm abs}^3 (\rm p)\exp(-\tau_{\rm abs})}\right\}^{1/(\rm 2p+13)} 
	\label{eqA8}
\end{equation}
With $B = \nu_{\rm abs}/(\gamma_{\rm abs}^2 \nu_{o})$, one finds from Equation (\ref{eqA7})
\begin{equation}
	B = \frac{3.6\times 10^3}{ \hat{\gamma}_{\rm abs}^2(\rm p)} \frac{\nu_{\rm abs,10}}{y^{4/(\rm 2p+13)} F_{\nu_{\rm abs},27}^{2/(\rm 2p+13)}},
	\label{eqA9}
\end{equation}
so that $\hat{B}^{\rm synch}(\rm p) = 3.6\times 10^3 / \hat{\gamma}_{\rm abs}^2(\rm p)$. Equations (\ref{eqA5}) and (\ref{eqA9}) can be combined to obtain an expression for the source radius
\begin{equation}
	R = \frac{2^4\times 10^{-10}}{\pi^{3/2}} c\hat{\gamma}^{\rm p+6}\frac{\tau_{\rm abs}(\rm p)}{\zeta(\rm p)} \frac{F_{\nu_{\rm abs},27}^{\rm (p+6)/(2p+13)}}
	{\nu_{\rm abs,10}\,y^{\rm 1/(2p+13)}}.
	\label{eqA10}
\end{equation}
A comparison with Equation (\ref{eq2.4}) shows that $\hat{R}(\rm p) = 8.6\times 10^{-15}  \hat{\gamma}^{\rm p+6}(\rm p) \tau_{\rm abs}(\rm p)/\zeta(\rm p)$.

The brightness temperature at the spectral peak can be written
\begin{equation}
	T_{\nu_{\rm abs}}^{\rm br,synch} = \frac{(\rm p+4)\chi({\rm p})\tau_{\rm abs}(\rm p)\exp(-\tau_{\rm abs})}{15\zeta(\rm p)}\frac{\gamma_{ \rm abs}mc^2}{k},
	\label{eqA11}
\end{equation}
which yields $\hat{T}(\rm p) = 4.0 \times 10^{-2} (\rm p+4)\chi(\rm p)\hat{\gamma}_{\rm abs}(\rm p)\exp(-\tau_{\rm abs}) \tau_{\rm abs}(\rm p)/\zeta(\rm p)$.
Furthermore, with $\gamma_{\rm abs}^2 = \nu_{\rm abs}/\nu_{o} B$ and $T_{\nu_{\rm abs}}^{\rm br,synch} = T_{\nu_{\rm abs}}^{\rm br,obs}$, one finds from Equation (\ref{eqA11})
\begin{equation}
	B = \left\{\frac{(\rm p+4)\chi(\rm p)\tau_{\rm abs}(\rm p)\exp(-\tau_{\rm abs})}{15 \zeta(\rm p)}\right\}^2 \left(\frac{mc^2}{k}\right)^2
	\frac{\nu_{\rm abs}}{\nu_{o} \left(T_{\nu_{\rm abs}}^{\rm br,obs}\right)^2}.
	\label{eqA12}
\end{equation}
 The alternative normalisation of $B$ is then $\hat{B}^{\rm obs}(\rm p) = 5.6 \times \{(\rm p+4) \chi(\rm p) \exp(-\tau_{\rm abs}) \tau_{\rm abs}(\rm p)/\zeta(\rm p)\}^2$.


\clearpage

\begin{thebibliography}{}
     
    \bibitem[Bartel et al.(2002)]{bar02} Bartel, N., Bietenholz, M.F., Rupen, M.P., et al., 2002, \apj, 581, 404
    
     \bibitem[Benvenuto et al.(2013)]{ben13} Benvenuto, O.G., Bersten, M.C., \& Nomoto, K., 2013, \apj, 762, 74
    
    \bibitem[Bersten et al.(2012)]{ber12} Bersten, M.C., Benvenuto, O.G., Nomoto, K., et al. 2012, \apj, 757, 31
    
    \bibitem[Bersten et al.(2018)]{ber18} Bersten, M.C., Folatelli,G., Garcia,F., et al. 2018, \nat, 554, 497
    
    \bibitem[Bietenholz et al.(2003)]{bie03} Bietenholz, M.F., Bartel, N., \& Rupen, M.P., 2003, \apj, 597, 374
    
    \bibitem[Bj\"{o}rnsson(2013)]{bjo13} Bj\"{o}rnsson, C.-I., 2013, \apj, 769, 65
    
    \bibitem[Bj\"{o}rnsson(2015)]{bjo15} Bj\"{o}rnsson, C.-I., 2015, \apj, 813, 43
    
    \bibitem[Bj\"{o}rnsson(2021)]{bjo21} Bj\"{o}rnsson, C.-I., 2021, \apj, 923, 61
    
    \bibitem[Bj\"{o}rnsson(2022)]{bjo22} Bj\"{o}rnsson, C.-I., 2022, \apj, 936, 98
    
    \bibitem[Bj\"{o}rnsson(2024)]{bjo24} Bj\"{o}rnsson, C.-I., 2024, \apj, 963, 93
    
    \bibitem[Bj\"{o}rnsson(2025)]{bjo25} Bj\"{o}rnsson, C.-I., 2025, \apj, 987, 3
    
    \bibitem[Bj\"{o}rnsson \& Keshavarzi(2017)]{b/k17} Bj\"{o}rnsson, C.-I., \& Keshavarzi, S.T., 2017, \apj, 841, 12
    
    \bibitem[Blandford \& K\"{o}nigl(1979)]{b/k79} Blandford, R.D., \& K\"{o}nigl, A., 1979, \apj, 232, 34
    
    \bibitem[Chevalier(1982a)]{che82a} Chevalier, R.A., 1982a, \apj, 258, 790
    
    \bibitem[Chevalier(1982b)]{che82b} Chevalier, R.A., 1982b, \apj, 259, 302
    
    \bibitem[Chevalier \& Fransson(1994)]{c/f94} Chevalier, R.A., \& Fransson, C., 1994, \apj, 420, 268
    
    \bibitem[Chevalier \& Soderberg(2010)]{c/s10} Chevalier, R.A., \& Soderberg, A.M., 2010, \apj, 711, L40
    
    \bibitem[Cotton et al.(1980)]{cot80} Cotton, W.D., Wittels, J.J., Shapiro, I.I., et al., 1980, \apj, 238, L123
    
    \bibitem[Dessart et al.(2018)]{des18} Dessart, L., Yoon, S.-C., Livne, E., \& Waldman, R., 2018, \aap, 612, 61
    
    \bibitem[de Witt et al.(2016)]{dew16} de Witt, A., Bietenholz, M.F., Kamble, A., et al., 2016, \mnras, 455, 511
    
    \bibitem[Ergon et al.(2014)]{erg14} Ergon, M., Sollerman, J., Fraser, M., et al., 2014 \aap, 562, 17
    
    \bibitem[Farah et al.(2025a)]{far25a} Farah, J.R., Howell, A.M., Terreran, G., et al. 2025a, \apj, 984, 60
    
    \bibitem[Farah et al.(2025b)]{far25b} Farah, J.R., Prust, L.J., Terreran, G., et al. 2025b, arXiv:2509.20601
    
    \bibitem[Filippenko(1997)]{fil97} Filippenko, A.V., 1997, \araa, 35, 309
    
    \bibitem[Fransson \& Bj\"{o}rnsson(1998)]{f/b98} Fransson, C., \& Bj\"{o}rnsson, C.-I., 1998, \apj, 509, 861
    
    \bibitem[Hamilton \& Sarazin(1984)]{h/s84} Hamilton, A.J.S., \& Sarazin, C.L., 1984, \apj, 281, 682
    
    \bibitem[Horesh et al.(2013)]{hor13} Horesh, A., Stockdale, C., Fox, D.B., et al., 2013, \mnras, 436, 1258
    
    \bibitem[Jerkstrand et al.(2015)]{jer15} Jerkstrand, A., Ergon, M., Smartt, S.J., et al., 2015, \aap, 573, A12
    
    \bibitem[Krauss et al.(2012)]{kra12} Krauss, M.I., Soderberg, A.M., Chomiuk, L., et al., 2012, \apj, 750, L40
    
    \bibitem[Marcaide et al.(2009)]{mar09} Marcaide, J.M., Mart\'{i}-Vidal, I., Alberdi, A., et al., 2009, \aap, 505, 927
    
    \bibitem[Marion et al.(2014)]{mar14} Marion, G.H., Vinko, J., Kirshner, R.P., et al., 2014, \apj, 781, 69
    
    \bibitem[Mart\'{i}-Vidal et al.(2024)]{mar24} Mart\'{i}-Vidal, I., Bj\"{o}rnsson, C.-I., P\'{e}rez Torres, M.A.,
     Lundqvist, P., \& Marcaide, J.M., 2024, \aap, 691, A171
    
    \bibitem[Nadyozhin(1985)]{nad85} Nadyozhin, D.K., 1985, \apss, 112, 225
    
    \bibitem[Podsiadlowski et al.(1993)]{pod93} Podsiadlowski, Ph., Hsu, J.J.L., Joss, P.C., et al., 1993, \nat, 364, 509
    
    \bibitem[Smartt(2015)]{sma15} Smartt, S.J., 2015, \pasa, 32, e016
    
    \bibitem[Soderberg et al.(2012)]{sod12} Soderberg, A.M., Margutti, R., Zauderer, B.A., et al., 2012, \apj, 752, 78
    
    \bibitem[Truelove \& McKee(1999)]{t/m99} Truelove, J.K., \& McKee, C.F., 1999, \apjs, 120, 299
    
    \bibitem[van der Laan(1966)]{van66} van der Laan, H., 1966, \nat, 211, 1131
    
    \bibitem[Van Dyk et al.(2013)]{van13} Van Dyk, S.D., Zheng, W., Clubb, K.I., et al., 2013, \apj, 772, L32
    
    \bibitem[Van Dyk et al.(2023)]{van23} Van Dyk, S.D., de Graw, A., Baer-Way, R., et al., 2023, \mnras, 519, 471
    
    \bibitem[Venkattu et al.(2023)]{ven23} Venkattu, D., Lundqvist, P., P\'{e}rez Torres, M., et al., 2023, \apj, 953, 157

    \bibitem[Weiler et al.(2007)]{wei07} Weiler, K.W., Williams, C.L., Panagia, N., et al. 2007, \apj, 671, 1959
    
    \bibitem[Woosley et al.(1994)]{woo94} Woosley, S.E., Eastman, R.G., Weaver, T.A., \& Pinto, P.A., 1994, \apj, 429, 300
    
    \bibitem[Yadav et al.(2016)]{yad16} Yadav, N., Ray, A., \& Chakraborti, S., 2016, \mnras, 459, 595

    
    
       
\end{thebibliography}
\end{document}